\documentclass[journal]{IEEEtran} 
\usepackage{graphicx} 
\usepackage{subfigure}
\usepackage{epstopdf}
\usepackage{amsmath,amssymb}
\usepackage{booktabs}
\usepackage{natbib}
\setcitestyle{square,sort,comma,numbers}
 
\begin{document}
\title{Robust Individual Circadian Parameter Estimation for Biosignal-based
Personalisation of Cancer Chronotherapy} 

\author{B.~Schott, J.~Stegmaier, A.~Arbaud, M.~Reischl, R.~Mikut
and~F.~L\'{e}vi
\thanks{B.S. is with the Institute for Applied Computer Science, Karlsruhe
Institute of Technology, Karlsruhe, Germany e-mail: benjamin.schott@kit.edu.}
\thanks{J.S. and M.R. are with the Institute
for Applied Computer Science, Karlsruhe Institute of Technology, Germany}%
\thanks{A.A. was with the Institut National de la Sanit\'{e} et de la Recherche
M\'{e}dicale, Paris}
\thanks{R.M. is Professor at the Institute for Applied Computer Science,
Karlsruhe Institute of Technology, Germany}
\thanks{F.L. is Professor at the Warwick Medical School, Warwick Systems Biology
Center, England and was former Professor at the Institut National de la
Sanit\'{e} et de la Recherche M\'{e}dicale, Paris}
}

\markboth{Proceedings of the WORKSHOP BIOSIGNAL PROCESSING 2016, April 7th\,-\,8th, 2016, Berlin, Germany}%
{Kurths \MakeLowercase{\textit{et al.}}: Please copy your title here}

\maketitle


\begin{abstract}
In cancer treatment, chemotherapy is administered according a constant schedule.
The chronotherapy approach, considering chronobiological drug delivery, adapts
the chemotherapy profile to the circadian rhythms of the human organism.
This reduces toxicity effects and at the same time enhances efficiency of
chemotherapy. 
To personalize cancer treatment, chemotherapy profiles have to be further
adapted to individual patients.
Therefore, we present a new model to represent cycle phenomena in circadian
rhythms. The model enables a more precise modelling of the underlying circadian
rhythms.
In comparison with the standard model, our model delivers better results in all
defined quality indices. 
The new model can be used to adapt the chemotherapy profile efficiently to
individual patients. The adaption to individual patients contributes to the aim
of personalizing cancer therapy.
\end{abstract}


\IEEEpeerreviewmaketitle

\section{Introduction}
%

Cancer is a major public health problem all over the world. It is the
leading cause of death in economically developed countries and the second
leading cause in developing countries.
There is also a big economic impact of cancer through a huge amount of
financial costs for both the person suffering from cancer and the society.
The general aim is to increase the overall survival rate.
Therefore, the treatment efficiency for patients suffering from cancer has to be
improved \cite{Levi2002}.



%

For this purpose circadian drug delivery in chemotherapy is
used.
Here, the tolerability and
efficiency of anticancer drugs are improved on the basis of the circadian timing
system in humans \cite{Levi2010}. Today, often the standard chemotherapy as well
as the radiation dose are chosen individually according to the tumour type. The new approach,
considering the circadian drug delivery, further adapts the chemotherapy profile
to circadian rhythms (chronomodulation) of the human organism. This reduces
toxicity effects of the chemotherapeutic drugs and at the same time enhances the
efficiency.
To access circadian rhythms \cite{Sarabia2008} in human, previous work dealt
with the requirements to measure a marker rhythm which characterizes the timing
of the internal circadian system \cite{Levi2007_1}. Such a marker rhythm has to be
periodic and easy to measure over a long period of time with non-invasive methods.

  
Up to now, the state-of-the-art was a fixed chronomodulated chemotherapy
profile for all patients \cite{Levi2003}. This fixed profile was
extracted from experimental models of the rest-activity rhythm \cite{Innominato2010,Levi2010}.
To personalize profiles, information out of biosignals (e.g., rest-activity,
temperature) was required to access the patient-specific circadian rhythms.
Such biosignals containing cycle phenomena are often modelled using the
Cosinor model \cite{Innominato2009_2,Mormont2000}. To model the rhythms, a
cosine wave is fitted to the data using optimization algorithms. In general, the
parameters of a cosine wave are used to describe the underlying rhythm \cite{Scully2011}.
However, the Cosinor model is not able
to model different lengths of intervals within one period of a cycle phenomenon
\cite{Sarabia2008}. 
%
%

Therefore, the estimation of such rhythms containing different lengths
of intervals within one period of a cycle phenomenon is not adequate using the
Cosinor analysis \cite{Ballesta2012}. This results in an increasing need for
novel models considering the different lengths of intervals \cite{Billy2012,
Clairambault2011}. This leads to a better description of individual differences
in individual patients to further optimize the benefit of chronomodulated chemotherapy \cite{Innominato2010, Levi2010}.
The analysis of such models and the extraction of relevant information to
characterize the patients is an important topic in research
\cite{Gachon2011, Levi2006, Levi2011, Mikut2008}.

To achieve this aim we developed a new method called Extended Cosinor model
\cite{Schott14}.
This model is able to represent cycle phenomena containing different lengths of
intervals within one period.
Therefore, a new segment-wise defined function is developed consisting of two
cosine waves. With the different circular frequencies of the two cosine waves it
is possible to model the different lengths of intervals. Also through the
segment-wise definition of the model function it is possible to model the
superior periodicity of the underlying rhythm.

%

This more detailed model of cycle phenomena in biosignals can be used to adapt
the chronomodulated chemotherapy efficiently to individual patients. This
contributes to the aim of personalizing cancer therapy.

%
 
\section{Methods}

\subsection{Extended Cosinor model}

\begin{figure}[!h]
	\centering
	\includegraphics[width=0.5\textwidth]{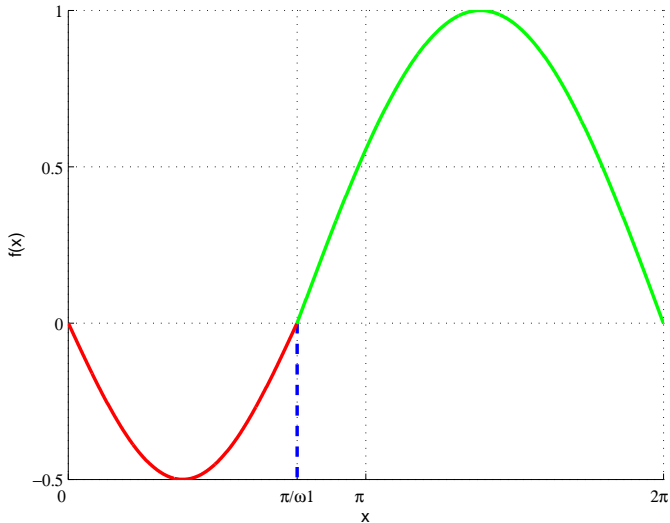}
	\caption[Model function for the Extended Cosinor]{Model function of the
	Extended Cosinor model: The two shifted cosine half-waves are shown in red
	($f_{cos1}$) and green ($f_{cos2}$). Due to the fact that $\varphi_1$ is
	equal to zero in this case, the intersection is at $\pi/\omega1$}
	\label{fig:Example of the new function}
\end{figure}

For the representation of the anti-periodicity within rhythms the Extended
Cosinor model was developed.
The model is able to consider different lengths of two intervals (e.g. day and
night, rest and activity) within one period. At the same time the model is able
to consider the periodicity of the superior rhythm (e.g. 24-hour cycle).
As a basis for the Extended Cosinor model, a new segment-wise defined model
function is developed ($f_\text{cos,Extended}$).
This model function consists of two cosine half-waves. The circular frequencies
of the two cosine waves are used to model the different lengths of the intervals
within one period. See Figure \ref{fig:Example of the new function} for
an example of the newly developed model function, where two cosine waves were
combined to a new function.

\subsection{Using the Extended Cosinor as a model for rhythms}
\begin{figure}[!htb]
	\centering
	\includegraphics[width=0.5\textwidth]{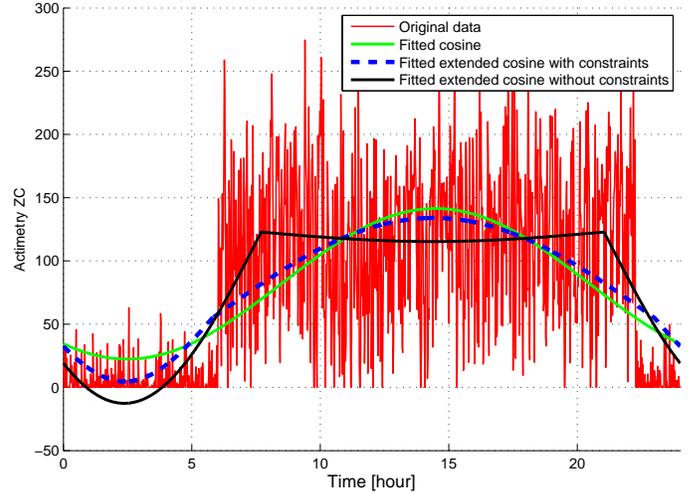}
	\caption[Different methods to fit a function to the rest-activity
	data]{Different methods to fit a model function to the biosignal (in this
	case a rest-activity rhythm). Here the actimetry in the zero crossing mode
	(ZC) is shown over the span of a whole day.}
	\label{fig:Different methods to fit the data}
\end{figure}

For modelling rhythms with the Extended Cosinor model, the model function
$f_\text{cos,Extended}$ has to be fitted to the original data extracted from the
measured biosignals.
For this purpose, a numerical optimization algorithm was used to estimate the
parameter vector of the model function. Therefore, a quality index describing
the goodness of the model was used to optimize these parameters.
Furthermore, constraints are defined within the optimization algorithm. 
These constraints force the parameters of the model function to be set in a
predefined interval. These intervals influence the shape of the function
following characteristic patterns. Therefore, the constraints within the optimization
algorithm are required for a better representation of the patterns within
a rhythm (see Figure \ref{fig:Different methods to fit the data}). All models
and algorithms used were implemented using the Matlab toolbox Gait-CAD
\cite{Mikut2008a}.

\section{Results}




For the evaluation of our model we used two databases (Database A and Database
B). 
Database A was created as part of a study in 2013 \cite{INSERM, Roche2014}. Two
biosignals, the temperature and the rest-activity signal, were recorded using
wireless sensors (sampling rate of $\frac{1}{min}$).
Both biosignals were monitored for 4 days in 10 patients suffering from cancer (for
more details see \cite{Schott14}).
Database B was created in connection with the PICADO project in 2012
\cite{PICADO, INSERM}. Here also the temperature and the rest-activity signal
were recorded in 34 healthy people over a period of 4 days (sampling rate of
$\frac{5}{min}$).
For evaluation purposes we used the rest-activity signal in both databases as an
example of a biosignal.
The Cosinor model, as the standard model for representing rest-activity rhythms
is compared with the new Extended Cosinor model using the quality indices listed
in Table \ref{tab:Quality indices for evaluation}.

\begin{table}[htbp]

    \caption[Quality indices for evaluation]{Quality indices for evaluation}%
    \label{tab:Quality indices for evaluation}
  \centering
  \begin{footnotesize}
    \begin{tabular}{|p{0.7cm}||p{1.5cm}||l||p{1.5cm}|}
    \hline
    Quality index & Description & Abbreviation & Values for a good model \\
    \hline
    $Q_1$ & Sum squared errors & SSE & Lower \\
    $Q_2$ & $R^2$ & R squared & Higher \\
    $Q_3$ & $adjusted R^2$ &  Adjusted R squared & Higher\\
    $Q_4$ & RMSE & Root mean squared error & Lower\\
    $Q_5$ & $r_\text{xy}$ & Correlation coefficient & Higher\\
    \hline
    \end{tabular}%
    \end{footnotesize}

\end{table}%


The results of modelling all rest-activity rhythms in Database A are listed
in Table \ref{tab:valuation of different methods for the rest-activity time
series in Database A}. Here, the bold black values state the best results for
each quality measure. As shown in the results the Extended Cosinor model is
superior to the Cosinor models in all quality indices. An example of one rest-activity
rhythm modelled with the Cosinor and Extended Cosinor model is shown in Figure
\ref{fig:Example of the cosinor and extended cosinor analysis of one data point
in Database A}.

\begin{table}[!htbp] 

\caption{Evaluation of different methods to represent the rest-activity time
series in Database A. Here bold values represent better results.}
\label{tab:valuation of different methods for the rest-activity time series in Database A}

\centering 
\begin{scriptsize}
\begin{tabular}{|p{1.8cm}||l||l|} 
\hline & Cosinor model & Extended Cosinor model\\
\hline
$Q_1$ Mean $\pm$ STD & 1.25e+07  $\pm$ 3.9e+06 & \bf{1.2e+07  $\pm$ 4e+06}\\
$Q_2$ Mean $\pm$ STD & 0.173  $\pm$ 0.122 & \bf{0.201  $\pm$ 0.144}\\
$Q_3$ Mean $\pm$ STD & 0.173  $\pm$ 0.122 & \bf{0.198  $\pm$ 0.145}\\
$Q_4$ Mean $\pm$ STD & 92.4  $\pm$ 13.8 & \bf{90.6  $\pm$ 13.7}\\
$Q_5$ Mean $\pm$ STD & 0.429  $\pm$ 0.172 & \bf{0.445  $\pm$ 0.164}\\
\hline
\end{tabular} 
\end{scriptsize}
 
\end{table}


For Database B, Table \ref{tab:Evaluation of different methods for the
rest-activity time series in Database B} lists the results of modelling the
rest-activity rhythm.
Here, also the Extended Cosinor model was superior to the Cosinor model
regarding all quality indices. Moreover, Figure \ref{fig:Example of the cosinor
and extended cosinor analysis of one data point in Database B} shows one example for
modelling a rest-activity rhythm in Database B.

\begin{table}[!htbp] 

\caption{Evaluation of different methods for the rest-activity time series in
Database B. Here bold values represent better results.}
\label{tab:Evaluation of different methods for the rest-activity time series in Database B} 

\centering 
\begin{scriptsize}
\begin{tabular}{|p{1.8cm}||l||l|} 
\hline & Cosinor model & Extended Cosinor model\\
\hline
$Q_1$ Mean $\pm$ STD & 2.82e+06  $\pm$ 2.92e+06 & \bf{2.67e+06  $\pm$ 2.65e+06}\\
$Q_2$ Mean $\pm$ STD & 0.339  $\pm$ 0.136 & \bf{0.369  $\pm$ 0.135}\\
$Q_3$ Mean $\pm$ STD & 0.338  $\pm$ 0.136 & \bf{0.367 $\pm$ 0.135}\\
$Q_4$ Mean $\pm$ STD & 37.6  $\pm$ 23.7 & \bf{36.7  $\pm$  23}\\
$Q_5$ Mean $\pm$ STD & 0.573  $\pm$ 0.124 & \bf{0.599  $\pm$ 0.115}\\

\hline
\end{tabular} 
\end{scriptsize}

\end{table}

The developed Extended Cosinor model provides better representation of rhythms
containing intervals with different lengths in comparison to the Cosinor model.
This advantage is based on the fact that the Extended Cosinor model in contrast
to the Cosinor model is able to model different lengths of intervals within one cycle
period. Furthermore, the underlying model function of the Extended Cosinor model
provides more parameters in comparison with the Cosinor model. The higher number
of parameters can be used for further accessing of the rhythms.

%
%

  \begin{figure}[!h]
	\centering
	\includegraphics[width=0.5\textwidth]{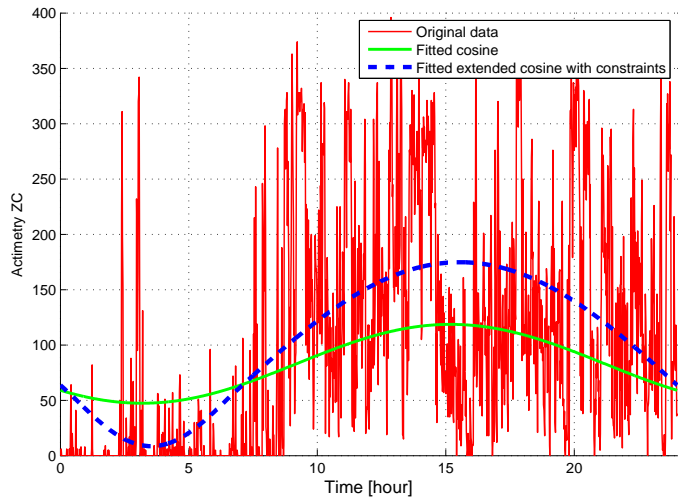}
	\caption[Example of the Cosinor and Extended Cosinor Analysis of one data point
	in Database A]{Example of the Cosinor and Extended Cosinor model for one
	rest-activity rhythm in Database A. Here it can be clearly observed, that the
	Extended Cosinor model (blue dashed line) models the data better in comparison
	with the Cosinor mode (green solid line).}
	\label{fig:Example of the cosinor and extended cosinor analysis of one data point in Database A}
\end{figure}
 
%
%
%
%
%

  \begin{figure}[!h]
	\centering
	\includegraphics[width=0.5\textwidth]{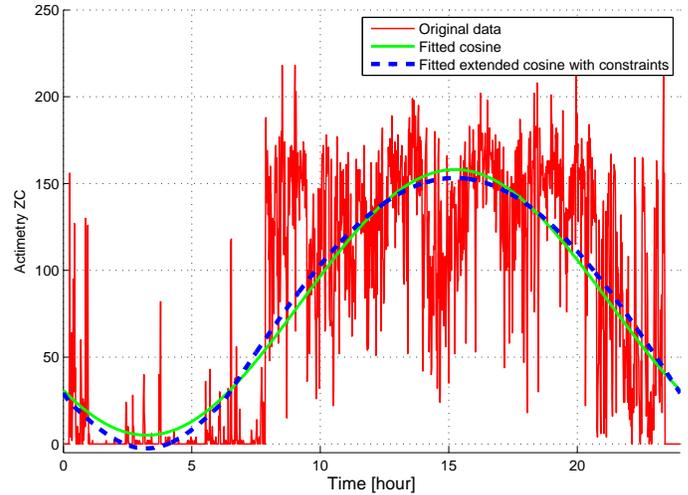}
	\caption[Example of the cosinor and extended cosinor analysis of one data point
	in Database B]{Example of the Cosinor and Extended Cosinor model for one
	rest-activity rhythm in Database B. Here it can be observed, that the Extended
	Cosinor model (blue dashed line) models the data better in comparison with the
	Cosinor mode (green solid line).}
	\label{fig:Example of the cosinor and extended cosinor analysis of one data point in Database B}
\end{figure}

%
%
%
\section{Conclusion}

In this paper a new model (Extended Cosinor) representing cycle phenomena in
rhythms is presented.
The new model is evaluated against the standard Cosinor model using
rest-activity rhythms as one exemplary biosignal.
The Extended Cosinor model provides a better representation of rhythms containing
intervals with different lengths within one cycle period.

%
%

%
%
%

The new model can be used to access biorhythms of individual
patients. In the context of chronomodulated chemotherapy, this can serve as a
basis for further personalization of chemotherapy according to the biorhythms of
the patients.
However, there are some limitations so far. If the biorhythms are too disrupted,
the quality of modeling is not as good as modelling normal rhythms. 
Therefore, future research has to handle the problem of adapting the constraints
for model generation to cope with such phenomena and further adapt these
constraints to every individual rhythm. It is suggested that adapting the
constraints improves the model quality for individual patients and therefore
makes a further steps towards the personalization of chemotherapy.

\section*{Acknowledgment}
The authors would like to thank the Institut National de la
Sani\'{e} et de la Recherche M\'{e}dicale (INSERM) for providing the datasets.

%

\ifCLASSOPTIONcaptionsoff
  \newpage
\fi



%
 
\bibliographystyle{abbrv}

\begin{thebibliography}{10}

\bibitem{Ballesta2012}
A.~Ballesta, J.~Clairambault, S.~Dulong, and F.~L\'{e}vi.
\newblock A Systems Biomedicine Approach for Chronotherapeutics Optimization:
  Focus on the Anticancer Drug Irinotecan.
\newblock In A.~Onofrio, P.~Cerrai, and A.~Gandolfi, editors, {\em New
  Challenges for Cancer Systems Biomedicine}, SIMAI Springer Series,
  301--327. Springer Milan, 2012.

\bibitem{Billy2012}
F.~Billy, J.~Clairambault, O.~Fercoq, T.~Lorenzi, A.~Lorz, and B.~Perthame.
\newblock Modelling Targets for Anticancer Drug Control Optimisation in
  Physiologically Structured Cell Population Models.
\newblock In {\em ICNAAM 2012 - 10th International Conference of Numerical
  Analysis and Applied Mathematics}, 1479: 1323--1326, Kos,
  Gr{\`e}ce, Sept. 2012. Simos, Thodoros, American Institute of Physics.

\bibitem{PICADO}
A.~Cis.
\newblock D\'{e}veloppement d'une Infrastructure Technologique Innovante et
  Polyvalente pour la Domom\'{e}decine.
\newblock http://www.utt.fr/fr/toute-l-actualite/domomedecine.html.

\bibitem{Clairambault2011}
J.~Clairambault, S.~Gaubert, and T.~Lepoutre.
\newblock Circadian Rhythm and Cell Population Growth.
\newblock {\em Mathematical and Computer Modelling}, 53: 1558 -- 1567, 2011.

\bibitem{Gachon2011}
F.~Gachon and D.~Firsov.
\newblock The Role of Circadian Timing System on Drug Metabolism and
  Detoxification.
\newblock {\em Expert Opinion on Drug Metabolism \& toxicology}, 7(2):147--158,
  2011.

\bibitem{Innominato2009_2}
P.~Innominato, M.~Mormont, T.~Rich, J.~Waterhouse, F.~L\'{e}vi, and
  G.~Bjarnason.
\newblock Circadian Disruption, Fatigue, and Anorexia Clustering in Advanced
  Cancer Patients: Implications for Innovative Therapeutic Approaches.
\newblock {\em Integrative Cancer Therapies}, 8(4):361--370, 2009.

\bibitem{Innominato2010}
P.~Innominato, F.~L\'{e}vi, and G.~Bjarnason.
\newblock Chronotherapy and the Molecular Clock: Clinical Implications in
  Oncology.
\newblock {\em Advanced Drug Delivery Reviews}, 62:979 -- 1001, 2010.

\bibitem{INSERM}
INSERM.
\newblock Int\'{e}r\^{e}t de l'actim\'{e}trie dans le Suivi des Maladies
  Chroniques.
\newblock 2013.

\bibitem{Levi2006}
F.~L\'{e}vi.
\newblock Chronotherapeutics: The Relevance of Timing in Cancer Therapy.
\newblock {\em Cancer Causes \& Control}, 17(4):611--621, 2006.

\bibitem{Levi2002}
F.~L\'{e}vi, E.~Filipski, V.~King, X.~Li, T.~Granda, M.~Mormont, X.~Liu,
  B.~Claustrat, and H.~Hastings.
\newblock Host Circadian Clock as a Control Point in Tumor Progression.
\newblock {\em Journal of the National Cancer Institute}, 94(9):690--697, 2002.

\bibitem{Levi2007_1}
F.~L\'{e}vi, C.~Focan, A.~Karaboue, V.~Valette, D.~Focan-Henrard, B.~Baron, F.~Kreutz, and S.~Giacchetti.
\newblock Implications of Circadian Clocks for the Rhythmic Delivery of Cancer
  Therapeutics.
\newblock {\em Advanced Drug Delivery Reviews}, 59:1015 -- 1035, 2007.

\bibitem{Levi2011}
F.~L\'{e}vi and A.~Okyar.
\newblock Circadian Clocks and Drug Delivery Systems: Impact and Opportunities
  in Chronotherapeutics.
\newblock {\em Expert Opinion on Drug Delivery}, 8(12):1535--1541, 2011.

\bibitem{Levi2010}
F.~L\'{e}vi, A.~Okyar, S.~Dulong, P.~Innominato, and C.~J.
\newblock Circadian Timing in Cancer Treatments.
\newblock {\em Annu. Rev. Pharmacol. Toxicol.}, 50:377--421, 2010.

\bibitem{Mikut2008}
R. Mikut.
\newblock {\em {Data Mining in der Medizin und Medizintechnik}},
\newblock Universit\"atsverlag Karlsruhe, 2008.

\bibitem{Mikut2008a}
R.~Mikut, O.~Burmeister, and M.~Reischl.
\newblock The Open Source {Matlab} Toolbox {Gait-CAD} and its Application to
  Bioelectric Signal Processing.
\newblock {\em {In Proc., DGBMT-Workshop Biosignalverarbeitung, Potsdam}}, 2008.

\bibitem{Levi2003}
M.~Mormont and F.~L\'{e}vi.
\newblock Cancer Chronotherapy: Principles, Applications, and Perspectives.
\newblock {\em Cancer}, 97(1):155--169, 2003.

\bibitem{Mormont2000}
M.~Mormont, J.~Waterhouse, P.~Bleuzen, S.~Giacchetti, A.~Jami, A.~Bogdan,
  J.~Lellouch, J.~Misset, Y.~Touitou, and F.~Levi.
\newblock Marked 24-h Rest-activity Rhythms are Associated with Better Quality
  of Life, Better Response, and Longer Survival in Patients with Metastatic
  Colorectal Cancer and Good Performance Status.
\newblock {\em Clinical Cancer Research}, 6:3038, 2000.

\bibitem{Roche2014}
V.~Roche, A.~Mohamad-Djafari, P.~Innominato, A.~Karaboue, and L.~F. Gorbach, A.
\newblock Thoracic Surface Temperature Rhythms as Circadian Biomarkers for
  Cancer Chronotherapy.
\newblock {\em Chronobiology International}, 31(3):409--420, 2014.

\bibitem{Sarabia2008}
J.~Sarabia, M.~Rol, P.~Mendiola, and J.~Madrid.
\newblock Circadian Rhythm of Wrist Temperature in Normal-living Subjects: A
  Candidate of new Index of the Circadian System.
\newblock {\em Physiology Behavior}, 95(4):570 -- 580, 2008.

\bibitem{Schott14}
B.~Schott.
\newblock Development of an Automated Framework for the Biosignal-based
  Personalisation of Cancer Chronotherapy.
\newblock Master's thesis, Karlsruhe Institute of Technology, 2014.

\bibitem{Scully2011}
C.~Scully, A.~Karaboue, J.~Liu, WM. AMD~Meyer, P.~Innominato, K.~Chon,
  A.~Gorbach, and F.~L\'{e}vi.
\newblock Skin Surface Temperature Rhythms as Potential Circadian Biomarkers
  for Personalized Chronotherapeutics in Cancer Patients.
\newblock {\em Interface Focus}, 1(1):48--60, 2011.

\end{thebibliography}

\end{document}